\documentclass[aps,prd,amsmath,floats,floatfix, twocolumn,
superscriptaddress,nofootinbib,showpacs]{revtex4-1}

\usepackage[colorlinks, pdfborder={0 0 0}, plainpages=false]{hyperref}
\usepackage{graphicx}
\usepackage{xspace}
\usepackage[usenames,dvipsnames]{color}
\usepackage{amssymb}
\usepackage[normalem]{ulem} 
\usepackage[caption=false]{subfig}
\usepackage{multirow}

\newcommand{\ba}{\overline}
\newcommand{\wh}{\widehat}
\newcommand{\h}{H}
\newcommand{\hdh}{H_{DH}}

\newcommand{\p}{\phi}
\newcommand{\ta}{\theta}

\newcommand{\Caltech}{\affiliation{Theoretical Astrophysics 350-17,
    California Institute of Technology, Pasadena, CA 91125, USA}}

\begin{document}

\title{
Constructing a boosted, spinning black hole in the damped harmonic gauge
}

\author{Vijay Varma} \Caltech
\author{Mark A. Scheel} \Caltech

\date{\today}

\begin{abstract}
The damped harmonic gauge is important for numerical relativity
computations based on the generalized harmonic formulation of
Einstein's equations, and is used to reduce coordinate
distortions near binary black hole mergers.
However, currently there is no prescription to
construct quasiequilibrium binary black hole initial data in this
gauge. Instead, initial data are typically constructed using a
superposition of two boosted analytic single black hole solutions
as free data in the solution of the constraint equations.  
Then, a smooth time-dependent gauge transformation is done early in the
evolution to move into the damped harmonic gauge.  Using this
strategy to produce initial data in damped harmonic gauge would require 
the solution of a single black hole in this gauge, which is not known
analytically. In this work we construct a single boosted, spinning, equilibrium black hole in damped harmonic coordinates as a regular time-independent
coordinate transformation from Kerr-Schild coordinates.
To do this, we derive and solve a set of four coupled, nonlinear,
elliptic equations for this transformation, with appropriate boundary
conditions. This solution can now be used in the construction of damped
harmonic initial data for binary black holes.
\end{abstract}


\maketitle

\section{Introduction}
\label{Sec:Introduction}

Gauge freedom is one of the most elegant features of general
relativity. Numerical relativity, however, inherently breaks this
freedom, since one picks a particular set of coordinates to
represent the solution on the computer.
Gauge choices are particularly important in numerical
relativity, since a poor gauge choice can lead to coordinate singularities.

Here we consider numerical relativity simulations that use
the generalized harmonic formulation of the Einstein equations
~\cite{Friedrich1985, Garfinkle2002, Pretorius2005a, Lindblom:2007}.
In this formalism, the coordinates $x^a$ obey
\begin{gather}
\nabla^c \nabla_c x^a = \h^a,
\label{Eq:GaugeCondition}
\end{gather}
where the gauge source function $\h^a$ is an arbitrarily chosen function of the 
coordinates and of the 4-metric $\psi_{ab}$, but not of the derivatives of the
4-metric. Here $\nabla_a$ is the covariant derivative operator compatible with
$\psi_{ab}$. The coordinates $x^a$ are treated as four
scalars in Eq.~(\ref{Eq:GaugeCondition}), so that one can
write $\nabla^c \nabla_c x^a = -\psi^{bc} \, {^{(4)}}\Gamma^a{}_{bc}$, where
$^{(4)}\Gamma^a{}_{bc}$ are the Christoffel symbols associated with
$\psi_{ab}$. Despite the considerable freedom allowed in the choice of $\h^a$, 
in practice it is not straightforward to choose an $\h^a$ that leads to coordinates 
without singularities or large distortions.

One gauge choice that has been particularly 
successful in the numerical evolution of binary black hole (BBH) mergers is
to choose $\h^a$ to satisfy the
damped harmonic gauge~\cite{Lindblom2009c,Szilagyi:2009qz,Choptuik:2009ww}, given by 
Eqs.~(\ref{Eq:HLowerindex}) below.  In damped harmonic gauge,
the spatial coordinates and the lapse function obey damped wave equations,
and the damping terms suppress spatial and temporal
coordinate distortions that grow large near merging
black hole horizons when using simpler gauge choices. Damped harmonic gauge
is a key ingredient in BBH simulations that use the generalized
harmonic formulation of Einstein's equations~\cite{Mroue:2013PRL}.

In this paper we are interested in combining damped harmonic
gauge with another property that is often desirable in BBH
simulations: initial data that is as close to equilibrium (in a
co-rotating frame) as possible.  If the initial data, including the
gauge degrees of freedom, are close to stationary in a co-rotating
frame, then the subsequent evolution will be slowly-varying in this
frame (at least during the inspiral phase), leading to higher
accuracy and lower computational cost. However, there is currently
no good prescription for constructing BBH initial data that satisfy
both the properties of quasiequilibrium and of damped harmonic
gauge.

To further motivate the desire for BBH simulations that share both of these
properties, consider in more detail the construction of initial data for BBH
simulations using the code SpEC~\cite{SpecWebsite}, which we use as an example
in this paper. Initial data are constructed~\cite{Lovelace2008} using the
Extended Conformal Thin Sandwich (XCTS) \cite{York1999, Pfeiffer2003b} 
formalism, which is a reformulation of the Einstein constraint equations. The
free data in this formalism are the conformal 3-metric $\tilde{g}_{ij}$, the
trace of the extrinsic curvature $K$, and the initial time derivatives of these 
quantities $\partial_t \tilde{g}_{ij}$ and $\partial_t K$. These time
derivatives are customarily set to zero in a co-rotating frame; this is meant 
as a quasi-equilibrium condition. The other free data, $\tilde{g}_{ij}$ and $K$,
are constructed by superposing the analytic expressions for the (non-conformal) 
three-metric $g_{ij}$ and $K$ of two single black holes (BHs) in 
Kerr-Schild~\cite{MTW, Visser:2007fj} coordinates. With this choice of free
data, the XCTS equations are solved to yield a constraint satisfying initial
data set.

The generalized harmonic evolution equations require as initial data
the initial values and time derivatives of all components of the 4-metric.
The solution of the XCTS equations determines all of these except
for the initial time 
derivatives $\partial_t N$ and $\partial_t N^i$ of the lapse $N$ and shift
$N^i$. These initial time derivatives are customarily chosen to be zero in a
co-rotating frame at $t=0$; these are additional quasi-equilibrium conditions
meant to reduce initial gauge dynamics.  By rewriting the Christoffel symbols
in Eq.~(\ref{Eq:GaugeCondition}) in terms of time derivatives of the lapse and
shift, these quasi-equilibrium conditions can be written as conditions on
$\h^0$ and $\h^i$:
\begin{align}
\label{Eq:dtLapse}
0 &= \partial_t N   = N^j \partial_{j}N -N^2 K  +N^3 \h^0 , \\
0 &= \partial_t N^i = N^j \partial_{j}N^i -N^2 g^{ij} \partial_{j}(\log N)
+N^2 \Gamma^i \nonumber \\
&\qquad\qquad+ N^2 (\h^i+ N^i \h^0) .
\label{Eq:dtShift}
\end{align}
Here $g_{ij}$ is the spatial metric, and $\Gamma^i$ is the Christoffel symbol
associated with $g_{ij}$. Note that  $\h^a$ thus constructed does not
necessarily satisfy the damped harmonic gauge condition.

The quasi-equilibrium initial $\h^a$ constructed above is typically used only
during the very early inspiral of the BBH system. Once the black holes approach 
each other, this choice of $\h^a$ leads to coordinate singularities. So early
in the evolution a time-dependent gauge transformation is done to gradually
change $\h^a$ from its initial quasiequilibrium value into damped harmonic
gauge. Unfortunately, this gauge transformation can lead to several
complications: (1) The early evolution of the BBH initial data described above
is typically discarded as it is contaminated by spurious transients generally
referred to as \emph{junk radiation}~\cite{Zhang:2013gda, Lovelace2009}. The
junk radiation is caused by several physical effects, such as the initial
ringdown of each BH to its correct equilibrium shape. The transformation to
damped harmonic gauge that begins near the start of the evolution introduces
gauge dynamics, making it difficult to separate the physical junk radiation
from gauge effects. (2) In full general relativity there is no analytic
expression for the orbital parameters of two compact objects that yields a
quasi-circular orbit. So to produce initial data describing a quasi-circular
binary, we use an iterative procedure~\cite{PhysRevD.83.104034} in which we
guess orbital parameters, evolve the binary for a few orbits, measure the
eccentricity from the (coordinate) trajectories of the BHs, and then compute
new lower-eccentricity orbital parameters for the next iteration.
This procedure occurs at early times while the gauge transformation
(which affects BH trajectories) is active,
and this might make it difficult to achieve a desired eccentricity.
(3) Typically, the evolution
becomes more computationally expensive during the gauge transition, because of
additional gauge dynamics that must be resolved. (4) It is difficult to start
simulations at close separations, because merger occurs so quickly that there
is not enough time to transition smoothly to damped harmonic gauge before
merger.

Therefore, there are several possible benefits in constructing BBH initial data 
that satisfy the damped harmonic gauge condition and are in quasi-equilibrium.
If one could construct a time-independent representation of a {\it single black 
hole} in damped harmonic coordinates, then one could construct
quasi-equilibrium damped harmonic BBH data by using a superposition of two 
single BHs in these coordinates, rather than in Kerr-Schild coordinates, as
free data in the XCTS system. This would produce quasi-equilibrium BBH
data that are nearly in damped 
harmonic gauge near each of the two black holes.
We know that a 
time-independent solution for a single BH in damped harmonic coordinates
exists, because this is the final state of the merged black hole in BBH
simulations done in the damped harmonic gauge. Unfortunately, the form of such
a single-BH solution is not known analytically.

In this work, we construct a numerical solution for a boosted, spinning single
BH in damped harmonic coordinates. This is done as a regular, time independent,
coordinate transformation from Kerr-Schild coordinates. We show that one needs
to solve a set of four coupled, nonlinear, elliptic equations for this
transformation. After imposing appropriate boundary conditions, we solve these 
equations numerically. Finally, we test our solution using a single BH
evolution: We evolve a single BH that starts in Kerr-Schild coordinates and
then transitions into the damped harmonic gauge. We show that the final steady
state of this evolution agrees with our solution for a single BH in damped
harmonic coordinates.

Given the single-BH coordinate representation presented here, one can
construct initial data for a binary BH in damped harmonic gauge by
superposing two such single BHs.  We discuss the binary case in a
separate work \cite{Varma:2018bbhid}, in which we construct, evolve,
and compare several BBH initial data sets (including those initially
in harmonic gauge and in damped harmonic gauge), and in which we also
introduce new boundary conditions for the XCTS equations.

The rest of the paper is organized as follows. Section~\ref{Sec:DhGauge} 
describes the damped harmonic gauge. In Sec~\ref{Sec:SingleBH}, we develop a
method to construct a boosted, spinning single BH in the damped harmonic gauge.
In Sec~\ref{Sec:SingleBHEv} we validate our solution using a single BH
evolution. Finally, in Sec~\ref{Sec:Conclusion} we provide some
concluding remarks. Throughout this paper we use geometric units with $G=c=1$.
We use Latin letters from the start of the alphabet $(a,b,c,\dots)$ for
spacetime indices and from the middle of the alphabet $(i,j,k,\dots)$ for
spatial indices. We use $\psi_{ab}$ for the spacetime metric, $g_{ab}$ for the
spatial metric, $N$ for the lapse and $N^i$ for the shift of the constant-$t$
hypersurfaces.

\section{Damped Harmonic Gauge}
\label{Sec:DhGauge}
In this section we describe the damped harmonic gauge in more detail.
But instead of immediately discussing the damped harmonic gauge, we start
first with the simpler case of the harmonic gauge, which is defined by the
condition that each coordinate satisfies the covariant scalar wave equation:
\begin{equation}
\label{Eq:HarmGauge}
\nabla^c \nabla_c x^a = 0.
\end{equation}
Harmonic coordinates are not unique: different coordinates can
satisfy Eq.~(\ref{Eq:HarmGauge}) but have different initial conditions and 
boundary values.

Harmonic coordinates have proven to be
extremely useful in analytic studies in general relativity
\cite{deDonder1921,Lanczos22,Choquet1952,fischer_marsden72,cook_scheel97},
but numerical simulations of BBH in this gauge tend to fail as they approach
the merger stage. One reason for these failures might be that Eq.~(\ref{Eq:HarmGauge}) 
does not sufficiently constrain the coordinates; for example it
admits dynamical wavelike solutions.
Since all physical fields in numerical relativity are expressed in terms of
the coordinates, an ideal gauge condition would eliminate these unwanted gauge dynamics.

The dynamical range available to harmonic coordinates can be reduced by adding a
damping term, resulting in the damped harmonic gauge \cite{Szilagyi:2009qz}:
\begin{gather}
\label{Eq:DhGaugeCondition}
\nabla^c \nabla_c x^a = \hdh^a, \\
\hdh^a = \mu_L~\log\left(\frac{\sqrt{g}}{N}\right)~t^a 
- \mu_S~N^{-1}~g^a_{i}~N^i.
\label{Eq:HLowerindex}
\end{gather}
Here $t^a$ is the future directed unit normal to constant-t hypersurfaces,
$g_{ab}=\psi_{ab}+t_at_b$
is the spatial metric of the constant-$t$ hypersurfaces,
$g$ is the determinant of this metric, $N$ is the lapse, $N^i$ is the shift,
and 
$\mu_L$ and $\mu_S$ are positive damping factors chosen as follows:
\begin{gather}
\label{Eq:Mu_definition}
\mu_S = \mu_L = \mu_0~\left[\log\left(\frac{\sqrt{g}}{N}\right) \right]^2,
\end{gather}
where
\begin{gather}
\mu_0 = f_0(t)~\exp\left(-a ~\frac{R^2}{w^2}\right).
\label{Eq:Mu_0_definition}
\end{gather}
Equation~(\ref{Eq:Mu_definition}) describes the dependence of the damping
factors on metric components, and Eq.~(\ref{Eq:Mu_0_definition}) describes
rolloff factors that are used to reduce damped harmonic gauge to harmonic
gauge far from the origin or at early times.
In Eq.~(\ref{Eq:Mu_0_definition}), $R$ is the Euclidean distance from the origin
and $w$ is a length scale which we choose to be $100 M$, where $M$ is the total
mass of the system.  The dimensionless constant $a$ is chosen to be $34.54$,
so that the Gaussian factor reaches a value of $10^{-15}$ at $R=w$.
Finally,
$f_0(t)$ is an optional smooth function of time that we include
if the evolution is meant to transition from a different gauge into damped
harmonic gauge; this function is zero before the transition
and unity afterwards.  The precise values of the constants $w$ and $a$
are not important for the success
of damped harmonic gauge in BBH simulations; any choice that
results in $\mu_0\sim 1$ near the black holes and $\mu_0 = 0$
near the outer boundary should suffice.

This choice of the gauge source function $\hdh^a$ has the following benefits
\cite{Szilagyi:2009qz}:
(1) The spatial coordinates $x^i$ satisfy a damped wave equation and are driven
towards solutions of the covariant spatial Laplace equation on a timescale of $1/\mu_S$.
This tends to reduce extraneous gauge dynamics when $1/\mu_S$
is chosen to be smaller than the characteristic physical timescale.
(2) Similarly, the lapse satisfies a damped wave equation with damping factor 
$\mu_L$~\cite{Lindblom2009c}.
(3) This gauge condition controls the growth of $\sqrt{g}/N$, which tends to blow up near 
black hole horizons near merger in simpler gauges like the harmonic gauge.
(4) The gauge source function $\hdh^a$ depends only on the coordinates and the spacetime
metric, but not on the derivatives of the metric. 
This means that this gauge condition preserves the principal part of the 
Einstein equations in the generalized harmonic formalism~\cite{Lindblom2006}, 
and hence preserves symmetric hyperbolicity.
Like harmonic coordinates, damped harmonic coordinates are not unique:
any initial coordinate choice can be evolved using
Eq.~(\ref{Eq:DhGaugeCondition}) and will satisfy the damped harmonic
condition.

\section{Boosted, spinning black hole in damped harmonic gauge}
\label{Sec:SingleBH}

First consider harmonic (not damped harmonic) coordinates. Although harmonic
coordinates are not unique,
there is a unique coordinate representation of a single boosted,
charged, spinning black hole that satisfies the harmonic coordinate
condition Eq.~(\ref{Eq:HarmGauge}), is time-independent, and is
regular at the event horizon.  This coordinate representation can be
determined analytically~\cite{cook_scheel97} by considering a regular
coordinate transformation from Kerr-Schild coordinates.

The situation is similar for damped harmonic coordinates.
In this section, we construct the unique coordinate representation of a
boosted, spinning single black hole that satisfies the damped harmonic
condition, Eqs.~(\ref{Eq:DhGaugeCondition})--(\ref{Eq:HLowerindex}), is 
time-independent, and is regular at the event horizon. Following 
Ref.~\cite{cook_scheel97}, we construct this solution by considering a 
coordinate transformation from Kerr-Schild coordinates. But unlike the case of
harmonic coordinates, for damped harmonic coordinates we will obtain a
numerical rather than an analytical solution.

Starting with Kerr-Schild coordinates (denoted by $x^{\ba{a}}$), we try to find
a transformation to new coordinates $x^a$ that satisfy the damped harmonic
condition,
\begin{gather}
\label{Eq:DhGaugeConditionLong}
\nabla^{c} \nabla_{c} x^{a} =
\frac{\partial_b \left(\sqrt{-\psi}~ \psi^{ab}\right)}{\sqrt{-\psi}} = \hdh^{a},
\end{gather}
where $\psi$ is the determinant of the spacetime metric $\psi_{ab}$.

For simplicity, we start with Kerr-Schild coordinates that represent 
an unboosted black hole. However, we
desire our damped harmonic coordinates to
represent a boosted black hole, so that we can use them in BBH initial data
where the two BHs are in orbit. To obtain a boosted BH we can apply a Lorentz
transformation. For fully harmonic coordinates (as opposed to damped harmonic
coordinates), adding a boost is not difficult, because applying a Lorentz
transformation to harmonic coordinates results in boosted coordinates that
still satisfy the harmonic gauge condition~\cite{cook_scheel97}.
However, this is not true for damped harmonic gauge. To see this, consider a
set of coordinates $x^{\wh{a}}$, related to $x^a$ by a Lorentz transformation:
\begin{gather}
x^a = \Lambda^{a}{}_{\wh{b}}~ x^{\wh{b}}.
\end{gather}
Because $\Lambda^{a}{}_{\wh{a}}$ has only constant components and its
determinant is unity, Eq.~(\ref{Eq:DhGaugeConditionLong}) is transformed into:
\begin{gather}
\frac{\partial_{\wh{b}} \left(\sqrt{-\wh{\psi}}~ \psi^{\wh{a}\wh{b}}\right)}{\sqrt{-\wh{\psi}}}
= \nabla^{\wh{c}} \nabla_{\wh{c}} x^{\wh{a}} = \Lambda^{\wh{a}}{}_{a} \hdh^{a}.
\label{eq:UnboostedDhGaugeConditionWithoutSpin}
\end{gather}
As $\hdh^a$ is not a tensor, 
$\hdh^{\wh{a}} \neq \Lambda^{\wh{a}}{}_{a} \hdh^{a}$,
so the transformed coordinates $x^{\wh{a}}$ do not satisfy the damped harmonic
condition. Therefore instead of
constructing unboosted damped harmonic coordinates and boosting the
coordinates afterwards,
we must build the boost into the coordinate
construction, by demanding that the transformed coordinates $x^{\wh{a}}$
satisfy Eq.~(\ref{eq:UnboostedDhGaugeConditionWithoutSpin}).

Similarly, we desire a BH solution with an arbitrary spin direction, 
but it is most straightforward to work with Kerr-Schild coordinates with spin
along the z-axis. In order to construct damped harmonic coordinates with
generic spins, we can apply an additional rotation transformation
$R^{\widetilde{b}}{}_{\wh{b}}$ to 
Eq.~(\ref{eq:UnboostedDhGaugeConditionWithoutSpin}).

Combining the boost and the rotation, the equation that must be
satisfied for the coordinates $x^a$ to obey the damped harmonic condition
and to have the desired boost and spin direction is
\begin{gather}
\frac{\partial_{\wh{b}} \left(\sqrt{-\wh{\psi}}~ \psi^{\wh{a}\wh{b}}\right)}{\sqrt{-\wh{\psi}}}
= \nabla^{\wh{c}} \nabla_{\wh{c}} x^{\wh{a}} = T^{\wh{a}}{}_{a}~ \hdh^{a},
\label{Eq:UnboostedDhGaugeCondition}
\end{gather}
where
\begin{gather}
\label{Eq:LorentzBoost}
x^a = T^{a}{}_{\wh{b}} ~ x^{\wh{b}}, \\
T^{a}{}_{\wh{b}} =  \Lambda^{a}{} _{\widetilde{b}}~ R^{\widetilde{b}}{}_{\wh{b}}.
\end{gather}
We proceed as follows: we start with unboosted Kerr-Schild coordinates 
$x^{\ba{a}}$ with spin in the z-direction and find a transformation to
intermediate coordinates $x^{\wh{a}}$ such that $x^{\wh{a}}$ satisfies the
condition Eq.~(\ref{Eq:UnboostedDhGaugeCondition}). This means that $x^a$,
related to $x^{\wh{a}}$ by Eq.~(\ref{Eq:LorentzBoost}), satisfies the damped
harmonic condition (Eq.~(\ref{Eq:DhGaugeCondition})), while having the desired
spin direction and boost with respect to $x^{\ba{a}}$.

\subsection{Transformation to damped harmonic gauge}
\label{Subsec:IntermediateCoords}

We define a time-independent transformation from the Kerr-Schild
coordinates $x^{\ba{a}}$ to intermediate coordinates $x^{\wh{a}}$ as follows:
\begin{equation}
\label{Eq:Transformation}
\begin{aligned}
& x^{\wh{0}} = x^{\ba{0}} + 2M\log\left(\frac{2M}{r-r_{-}}\right) + U^{\wh{0}}(x^{\ba{i}}) ,\\
& x^{\wh{1}} = x^{\ba{1}} -M \sin{\ta} \cos{\p} + U^{\wh{1}}(x^{\ba{i}}) ,\\
& x^{\wh{2}} = x^{\ba{2}} -M \sin{\ta} \sin{\p} + U^{\wh{2}}(x^{\ba{i}}) ,\\
& x^{\wh{3}} = x^{\ba{3}} -M \cos{\ta}          + U^{\wh{3}}(x^{\ba{i}}),
\end{aligned}
\end{equation}
where $M$ is the mass of the black hole, $r_{-}=M-\sqrt{M^2-a^2}$ is the radius
of the Cauchy horizon, $a$ is the Kerr spin parameter and $(r,\ta,\p)$ are the
spatial coordinates of the spherical coordinate version of the standard
Kerr-Schild coordinates~\cite{MTW}: 
\begin{gather}
r^2 = \frac{\sum_{\ba{i}=1}^{3}(x^{\ba{i}})^2 \!-\! a^2}{2} + 
\sqrt{\frac{\left(\sum_{\ba{i}=1}^{3}(x^{\ba{i}})^2 \!-\! a^2\right)^2}{4}  
\!+\! (a x^{\ba{3}})^2} \\
\cos{\ta} = \frac{x^{\ba{3}}}{r} \\
\cos{\p} = \frac{rx^{\ba{1}}+ax^{\ba{2}}}{(r^2 + a^2) \sin{\ta}}
\end{gather}

Using Eq.~(\ref{Eq:Transformation}), the left hand side of 
Eq.~(\ref{Eq:UnboostedDhGaugeCondition}) can be written in terms of the
Jacobian of the transformation
$J^{\wh{a}}{}_{\ba{a}} = \partial x^{\wh{a}}/\partial x^{\ba{a}}$:
\begin{equation}
\label{Eq:Jacobian}
\frac{\partial_{\wh{b}} \left(\sqrt{-\wh{\psi}}~ \psi^{\wh{a}\wh{b}}\right)}{\sqrt{-\wh{\psi}}}
= \partial_{\ba{b}} \left( J^{\wh{a}}{}_{\ba{a}}~\psi^{\ba{a}\ba{b}} \right).
\end{equation}
Note that the Jacobian depends on first derivatives of
$U^{\wh{a}}$, so this is a second-order
elliptic equation for $U^{\wh{a}}$.

\subsubsection{Elliptic equations}
\label{Subsubsec:EllipticEqs}
After substituting the explicit form of the
Kerr-Schild metric~\cite{MTW} $\psi^{\ba{a}\ba{b}}$ into
Eq.~(\ref{Eq:UnboostedDhGaugeCondition}), and using Eq.~(\ref{Eq:Jacobian}),
a lengthy but straightforward
computation yields:
\begin{gather}
\label{Eq:EllipticEqs}
\mathcal{L} U^{\wh{a}} = T^{\wh{a}}{}_{a} \hdh^{a} ,\\
\mathcal{L} = \frac{\partial_r (\Delta \partial_r)}{\rho^2}
+\frac{\partial_{\ta}(\sin{\ta} \partial_{\ta})}{\rho^2 \sin{\ta}}
+\frac{\partial_{\p}^2}{\rho^2 \sin^2{\ta}}
+\frac{2 a \partial_{r}\partial_{\p}}{\rho^2},
\label{Eq:EllipticOpetator}
\end{gather}
where $\mathcal{L}$ is a linear differential operator, $\Delta=r^2-2 M r + a^2$,
and $\rho^2 = r^2 + a^2 \cos^2{\ta}$.

On the right hand side of these equations, $\hdh^a$ is obtained from 
Eq.~(\ref{Eq:HLowerindex}):
\begin{gather}
\label{Eq:HUpperindex}
\hdh^0 = \frac{\mu_0}{N} \left[\log\left(\frac{\sqrt{-\psi}}{N^2}\right) \right]^3  ,\\
\hdh^i = \frac{-\mu_0 N^i}{N} \left[\log\left(\frac{\sqrt{-\psi}}{N^2}\right) \right]^2
\label{Eq:LapseShift}
\left[1 + \log\left(\frac{\sqrt{-\psi}}{N^2}\right) \right] ,
\end{gather}
where
\begin{gather}
N = \sqrt{\frac{1}{-\psi^{00}}} ,\\
N^i = N^2~\psi^{0i},  \\
\psi^{ab} =  T^{a}{}_{\wh{a}} ~T^{b}{}_{\wh{b}}
~J^{\wh{a}}{}_{\ba{a}}~J^{\wh{b}}{}_{\ba{b}}~ \psi^{\ba{a}\ba{b}},
\end{gather}
and $\psi$ is the determinant of $\psi_{ab}$.

Finally, following Eq.~(\ref{Eq:Mu_0_definition}), we get
\begin{gather}
\mu_0 = \exp\left(-a ~\frac{\sum_i x^i x^i}{w^2}\right), \\
x^i =  T^{i}{}_{\wh{a}} ~J^{\wh{a}}{}_{\ba{a}}~x^{\ba{a}}.
\end{gather}

Eqs.~\ref{Eq:EllipticEqs} are a set of four
coupled, nonlinear elliptic equations with three independent variables
$(r,\ta,\p)$. Note that the left hand side of Eq.~(\ref{Eq:EllipticEqs}) is
linear in the functions $U^{\wh{a}}$ and all the nonlinearities come from the
source function $\hdh^a$ as seen in 
Eqs.~(\ref{Eq:HUpperindex}) and (\ref{Eq:LapseShift}) (the functions
$U^{\wh{a}}$ appear in the Jacobians $J^{\wh{a}}{}_{\ba{a}}$).
For harmonic coordinates, as the gauge source function is zero, the
four equations are decoupled,
linear, and separable in the radial and polar coordinates~\cite{cook_scheel97}.
In the more general case of damped harmonic coordinates, obtaining an analytical
solution is very challenging because the equations are coupled and nonlinear.
Therefore, we solve these elliptic equations numerically, using a spectral
elliptic solver~\cite{Pfeiffer2003}.

It is interesting to note that the principal part of the elliptic equations
is entirely on the left hand side, as $\hdh^a$ has only up to first derivatives
of the functions $U^{\wh{a}}$ (in the form of the Jacobians). 
Hence, the principal part is the same as that for harmonic coordinates, derived 
in Ref.~\cite{cook_scheel97}.

\subsubsection{Boundary conditions}
\label{Subsubsec:BoundaryConditions}

Before we can solve the elliptic equations derived above, we need to impose suitable
boundary conditions. The elliptic equations have three independent variables
$(r,\ta,\p)$. We do not need to specify a boundary condition for $\ta$ and $\p$
as we use spherical harmonic basis functions for the angular part in the elliptic solver.
For the radial outer boundary condition, we impose asymptotic flatness.
Note that Eq.~(\ref{Eq:Transformation}) is equivalent to writing
$x^{\wh{a}} = x^{a}_h + U^{\wh{a}}$, where $x^{a}_h$ are the fully harmonic
coordinates of Ref.~\cite{cook_scheel97}.  Because $x^{a}_h$ are
already asymptotically flat, our boundary condition is
\footnote{In practice, the outer boundary is set at a radius $\sim 10^{15}$ times the mass of the BH.}
\begin{equation}
\label{Eq:EllipticBCOuter}
\left. U^{\wh{a}} \right |_{r\to\infty} = 0.
\end{equation}

For the boundary condition at the inner radial boundary, consider the elliptic
equations, Eqs.~(\ref{Eq:EllipticEqs}) and (\ref{Eq:EllipticOpetator}),
with the radial derivatives expanded,
\begin{gather}
\frac{\Delta \partial^2_r U^{\wh{a}} }{\rho^2}
+\frac{2 (r-M) \partial_r U^{\wh{a}} }{\rho^2}
+\frac{\partial_{\ta}(\sin{\ta} \partial_{\ta} U^{\wh{a}} )}{\rho^2 \sin{\ta}} \nonumber \\
+\frac{2 a \partial_{r}\partial_{\p} U^{\wh{a}} }{\rho^2}
+ \frac{\partial^2_{\p} U^{\wh{a}} }{\rho^2 \sin^2{\ta}}
= T^{\wh{a}}_{a} \hdh^{a}.
\label{Eq:EllipticEqsExpaned}
\end{gather}
Now, $\Delta=0$ at $r=r_{+} = M + \sqrt{M^2-a^2}$, the event horizon. Therefore, at $r=r_{+}$
the first term of Eq.~(\ref{Eq:EllipticEqsExpaned}) goes to zero
and the nature of the principal part changes. In order to ensure regularity of coordinates at
the event horizon we restrict the domain to $[r_{+}, \infty)$ and impose a
regularity boundary condition at $r_{+}$:
\begin{gather}
\frac{2 (r-M) \partial_r U^{\wh{a}} }{\rho^2}
+\frac{\partial_{\ta}(\sin{\ta} \partial_{\ta} U^{\wh{a}} )}{\rho^2 \sin{\ta}}
+ \frac{\partial^2_{\p} U^{\wh{a}} }{\rho^2 \sin^2{\ta}} \nonumber \\
+\frac{2 a \partial_{r}\partial_{\p} U^{\wh{a}} }{\rho^2}
= T^{\wh{a}}_{a} \hdh^{a} ~~\textrm{at}~~ r\to r_{+}.
\label{Eq:EllipticBC}
\end{gather}

\subsection{Convergence tests}

Having chosen suitable boundary conditions for the elliptic equations, we
solve them numerically using a spectral elliptic solver \cite{Pfeiffer2003}.
Our domain consists of $12$ concentric spherical shells extending from
the horizon $r_+$ to $10^{15} M$, distributed roughly exponentially in
radius. Each shell has the same number of angular collocation points and
approximately the same number of radial points. The number of collocation
points in each subdomain is set by specifying an error tolerance to our
adaptive mesh refinement (AMR) algorithm~\cite{Ossokine:2015yla, 
Szilagyi:2014fna}.

The elliptic solver yields a solution for the intermediate coordinates
$x^{\wh{a}}$, from which we obtain the damped harmonic coordinates $x^a$ using 
Eq.~(\ref{Eq:LorentzBoost}). To quantify how well the final coordinates $x^a$
actually satisfy the damped harmonic gauge condition 
(Eq.~(\ref{Eq:DhGaugeCondition})), we define normalized damped harmonic
constraints and constraint energy\footnote{Notice that for the denominator of
Eq.~(\ref{Eq:DhCa}) below, repeated indices are summed over \emph{after}
squaring the quantities, unlike the standard summation notation.}:

\begin{gather}
\label{Eq:DhCa}
\mathcal{C}^a_{DH} = \frac{\Vert \psi^{bc}\,{^{(4)}}\Gamma^a{}_{bc} + \hdh^a \Vert}
{\left\Vert 
\sqrt{\sum\limits_{a,b,c=0}^{3} \Big[ (\psi^{bc}\,{^{(4)}}\Gamma^a{}_{bc})^2 
+ (\hdh^a)^2 \Big] } ~\right \Vert}, \\ 
\mathcal{C}_{DH} = \sqrt{\sum_{a=0}^3 \mathcal{C}^a_{DH} \mathcal{C}^a_{DH}} \,,
\label{Eq:DhConstraintEnergy}
\end{gather}
where $\Vert . \Vert$ is the $L^2$ norm over the domain. 
The numerator of Eq.~(\ref{Eq:DhCa}) is zero if Eq.~(\ref{Eq:DhGaugeCondition})
is exactly satisfied, and the denominator of Eq.~(\ref{Eq:DhCa}) is chosen so
that a solution very far from damped harmonic gauge has $\mathcal{C}^a_{DH}$ of
order unity.

\begin{figure}[hbt]
\begin{center}
\includegraphics[scale=0.55]{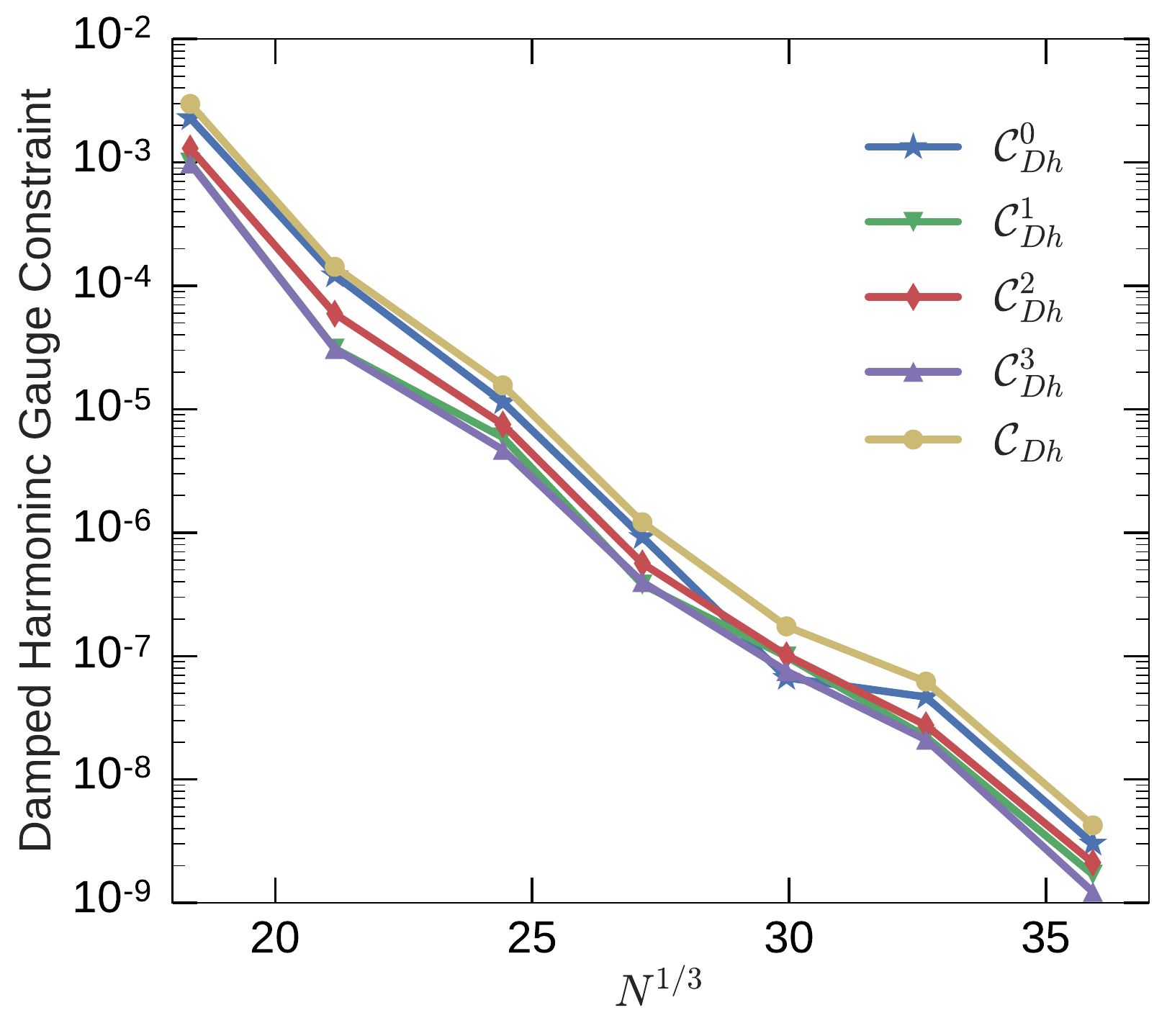}
\caption{Convergence test for solving the elliptic equations
(Eq.~(\ref{Eq:EllipticEqs})) to construct a single BH in the damped harmonic
gauge. Plotted are the damped harmonic constraints (cf. Eqs.~(\ref{Eq:DhCa})
and (\ref{Eq:DhConstraintEnergy})) as a function of the 
number of collocation points per dimension in the domain.
As expected for spectral methods, the
constraints converge exponentially.}
\label{Fig:ConvergenceTest}
\end{center}
\end{figure}

Figure~\ref{Fig:ConvergenceTest} shows the values of the damped harmonic
constraints as a function of numerical resolution, where
    higher resolution is achieved by setting a lower AMR error tolerance.
We note that the constraints 
decrease exponentially with resolution, as expected for a spectral method.

\begin{figure}[hbt]
\begin{center}
\includegraphics[scale=0.55]{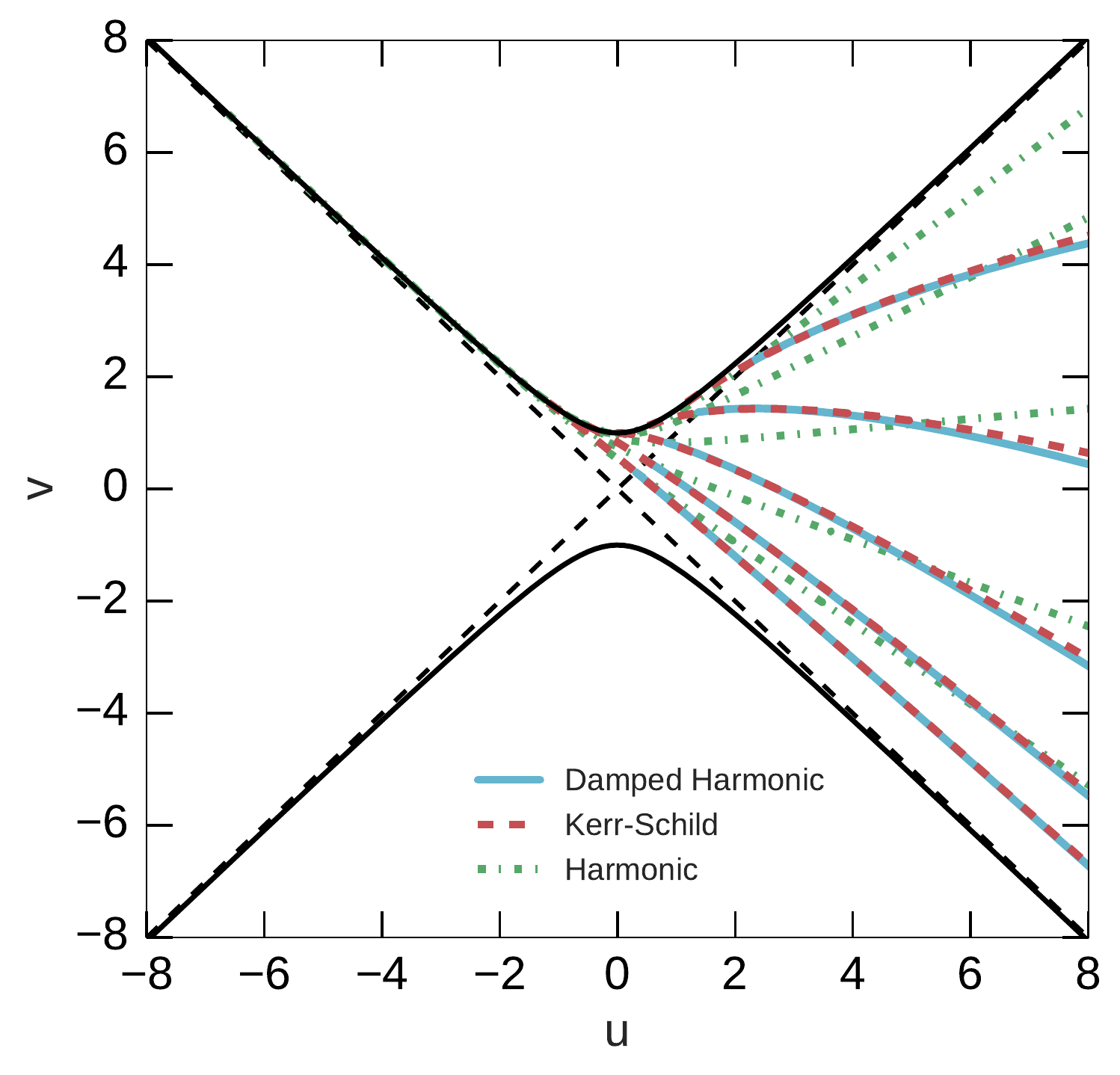}
\caption{Kruskal-Szekeres diagram showing constant time slices of the unique
horizon-penetrating time-independent slicings of Schwarzschild spacetime in
damped harmonic and harmonic coordinates, and constant time slices in 
Kerr-Schild coordinates. The solid black curves represent the curvature
singularity while the dashed black lines represent the event horizon. Note that 
the damped harmonic slices only extend up to the event horizon because we
restrict our numerical solution to this region; nevertheless, the damped
harmonic slices are horizon-penetrating. Interestingly, we see that the damped
harmonic slices are quite close to the Kerr-Schild slices.
}
\label{Fig:DhSlicing}
\end{center}
\end{figure}

\subsection{Choosing a time slice}

The solution of the elliptic equations along with Eq.~(\ref{Eq:LorentzBoost})
gives us a transformation from Kerr-Schild coordinates ($x^{\ba{a}}$) to
damped harmonic coordinates ($x^a$). But the desired initial data requires
computing the metric and its derivatives on a slice of constant time in the new 
coordinates $x^a$, so it is necessary to construct such a slice as a function
of the Kerr-Schild coordinates.
Using Eq.~(\ref{Eq:LorentzBoost}), we can construct a $x^0=0$ slice as follows:
\begin{gather}
x^0 = 0 = T^{0}{}_{\wh{a}}~ x^{\wh{a}}, \\
\label{Eq:DhTimeCoordTransform}
x^{\wh{0}} = \frac{-T^{0}{}_{\wh{i}}}{T^{0}{}_{\wh{0}}} ~x^{\wh{i}}, \\
x^i = T^{i}{}_{\wh{a}} ~x^{\wh{a}} =  \frac{-T^{i}{}_{\wh{0}} 
~T^{0}{}_{\wh{i}}}{T^{0}{}_{\wh{0}}} ~x^{\wh{i}} + T^{i}{}_{\wh{i}} ~x^{\wh{i}}.
\label{Eq:DhTimeSlice}
\end{gather}
This gives us a constant-time slice of damped harmonic coordinates ($x^0=0$, 
$x^i$) in terms of the intermediate coordinates ($x^{\wh{a}}$), which in turn
are expressed as a transformation from Kerr-Schild coordinates 
(Eqs.~(\ref{Eq:Transformation})).

The final step in constructing single-BH initial data is to
compute the metric and its derivatives on a slice of constant
$x^0=0$. This is done by choosing a set of points in the new
coordinates ($x^0=0$, $x^i$), computing the corresponding
$x^{\wh{a}}$ using Eqs.~(\ref{Eq:DhTimeCoordTransform})
and~(\ref{Eq:DhTimeSlice}), computing the corresponding Kerr-Schild
coordinates $x^{\ba{a}}$ using Eqs.~(\ref{Eq:Transformation}), and
evaluating the metric and its derivatives analytically
at those values of $x^{\ba{a}}$ using the Kerr-Schild expressions.
The components of the metric and its derivatives are then
transformed using the Jacobians (and Hessians for the metric derivatives)
that relate $x^{\ba{a}}$ and $x^a$.

To visualize the embedding of these damped harmonic slices in spacetime, we 
restrict ourselves to a nonspinning BH with zero boost. In this spherically
symmetric case, we can use the Kruskal-Szekeres coordinates to display the time 
slices on a spacetime diagram. These are shown in Fig~\ref{Fig:DhSlicing},
along with constant Kerr-Schild time slices and constant time slices of the
unique time-independent horizon-penetrating harmonic slicing of Schwarzschild
spacetime~\cite{cook_scheel97}. We note that constant time slices of damped
harmonic coordinates lie nearly on top of the constant time slices of 
Kerr-Schild coordinates, indicating that the extrinsic curvature of the two
slicings are quite similar.

\begin{figure*}[hbt]
\begin{center}
{\includegraphics[scale=0.6]{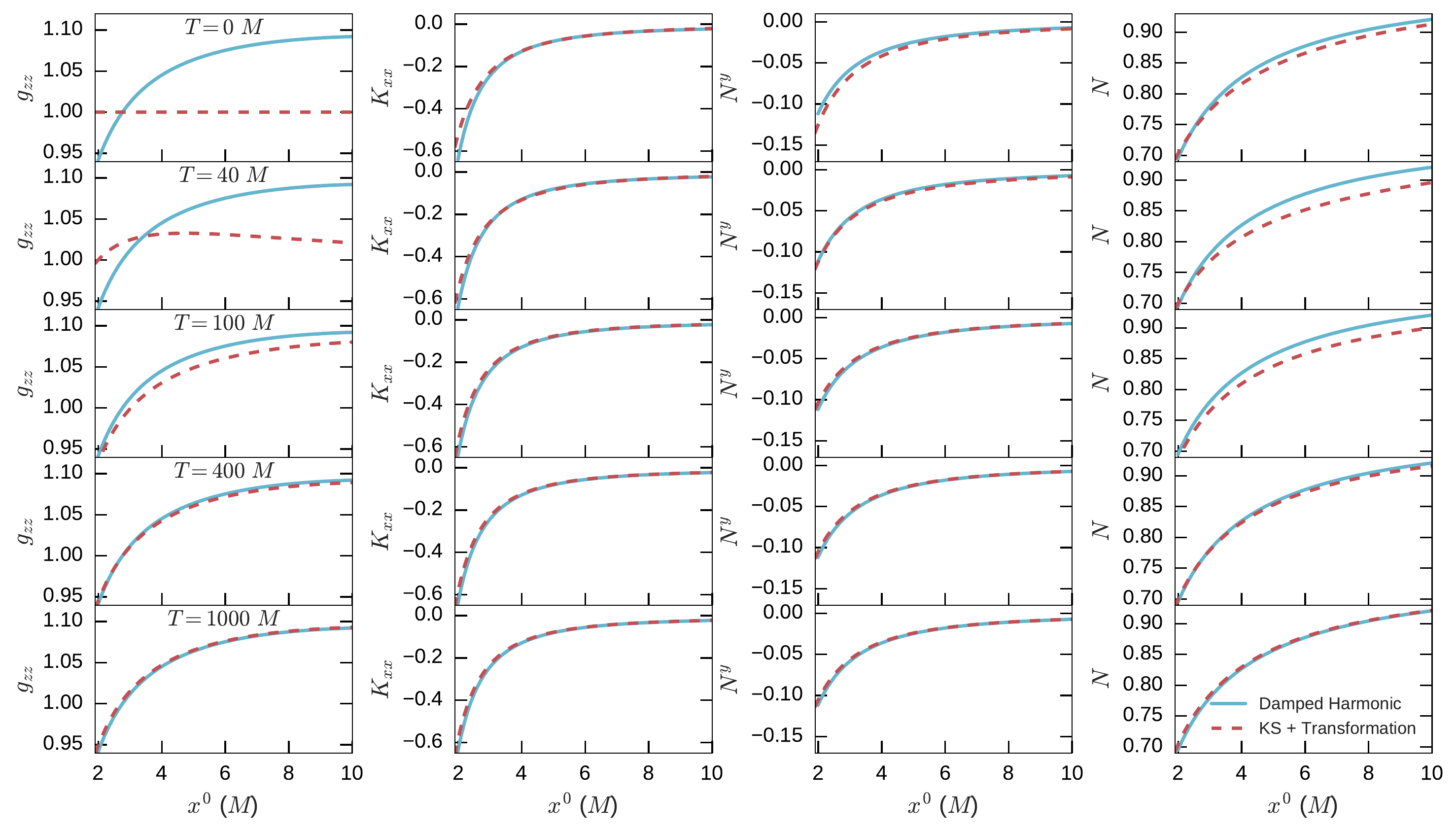}}
\caption{
Snapshots during the evolution of a single BH with mass $M=1$ and dimensionless
spin $\chi_z=0.5$, starting in Kerr-Schild coordinates and moving into damped
harmonic coordinates over a time scale of $50M$. Certain components of the
spatial metric $g_{ij}$, extrinsic curvature $K_{ij}$, shift $N^i$ and lapse
$N$ along the $x$-axis are shown as dashed red lines as the evolution
progresses. The horizontal axis shows the $x$ coordinate. For each column, time 
flows downwards and is shown in the left most column. The solid blue lines show 
our solution for a single time-independent BH in damped harmonic gauge, as
described in Sec.~\ref{Sec:SingleBH}. This solution agrees with the final state 
of the evolution.
}
\label{Fig:SBHEvMetric}
\end{center}
\end{figure*}

\section{Validation against single black hole simulations}
\label{Sec:SingleBHEv}

In this section, we check whether the solution we constructed in
Sec.~\ref{Sec:SingleBH} agrees with the time-independent final state
of a single BH that begins in some different gauge and is evolved
numerically using damped harmonic gauge conditions.

We start with a single BH on a $t=0$ slice of Kerr-Schild coordinates, and
we evolve it using the following time-dependent gauge source function:
\begin{gather}
 \h^a (t) = \wh{\h}^a ~ e^{-t^4/\sigma^4} + \hdh^a.\label{eq:GaugeSourceWithRolloff}
\end{gather}
Here $\wh{\h}^a$ is the equilibrium gauge source function satisfying
Eqs.~(\ref{Eq:dtLapse}) and~(\ref{Eq:dtShift}) for a single Kerr black hole
in Kerr-Schild coordinates. It is computed analytically 
as a known function of $t$ and $x^i$ during the evolution. 
$\hdh^a$ is the damped harmonic gauge source function given by 
Eq.~(\ref{Eq:HLowerindex}) and Eq.~(\ref{Eq:Mu_definition}), where we set 
$f_0(t) = 1 - e^{-t^4/\sigma^4}$. 
During the evolution, $\h_a$ is computed numerically using live values of 
the metric and its derivatives. We choose the time scale of the gauge transformation, 
$\sigma$, to be $50 M$. At early times, the BH remains time-independent in
Kerr-Schild coordinates, then there is a transition on a timescale of $50M$ in
which the solution is dominated by gauge dynamics, and at late times the
solution obeys the damped harmonic gauge condition and settles down to a 
time-independent state.

Figure \ref{Fig:SBHEvMetric} shows the evolution of certain components of the
metric as the evolution progresses. These are compared against
the single BH damped harmonic solution of Sec.~\ref{Sec:SingleBH}.
The final steady state solution of the simulation agrees with our solution for
the time-independent single BH in damped harmonic coordinates.
We note that the extrinsic curvature, lapse and shift of the initial state,
which is a black hole in Kerr-Schild coordinates, are quite close to
the corresponding quantities in the final state; these are all quantities
that depend on the embedding of the constant time hypersurfaces in spacetime.
We have already seen from Fig.~\ref{Fig:DhSlicing} that for zero spin, this
embedding is very similar for Kerr-Schild and damped harmonic slicings;
Fig.~\ref{Fig:SBHEvMetric} suggests that this embedding is also similar for
nonzero spin.

\section{Conclusion}
\label{Sec:Conclusion}

The damped harmonic gauge has been useful for simulations of binary black hole 
spacetimes, and is a key ingredient for handling mergers in simulations that use 
the generalized harmonic formalism. However, currently there is no prescription to 
construct quasi-equilibrium binary black hole initial data in this gauge; until now, 
there has been no prescription to construct even a time-independent {\it single}
black 
hole in this gauge.

In this work we have developed a method to construct a
time-independent boosted, spinning single black hole in damped
harmonic gauge.  We start with a black hole in Kerr-Schild
coordinates, and we construct a coordinate transformation to damped
harmonic coordinates. This transformation involves the numerical
solution of four coupled, nonlinear elliptic equations with
appropriate boundary conditions. We solve these equations with a
spectral elliptic solver, and we verify that the solution agrees with
the final time-independent state of a single black hole that begins in
Kerr-Schild coordinates and is evolved using the damped harmonic gauge.

Our procedure to construct a time-independent boosted, spinning, single BH in damped 
harmonic coordinates can now be used to construct equilibrium BBH initial data that 
satisfies the damped harmonic gauge. This is done by superposing two time-independent 
damped-harmonic BH solutions, in the same way that BBH initial data is currently built by 
superposing two time-independent Kerr-Schild BH solutions.

The next step is to use the solutions here to construct a BBH initial data set in damped 
harmonic gauge, evolve it, and compare with evolutions of BBH initial data sets in harmonic 
gauge and in superposed Kerr-Schild coordinates.  This is done in a separate 
work, Ref.~\cite{Varma:2018bbhid}.

\begin{acknowledgments}
This work was supported in part by the Sherman Fairchild Foundation
and NSF grants PHY-1404569, PHY-1708212, and PHY-1708213 at Caltech. The simulations were performed on the Wheeler cluster at Caltech,
which is supported by the Sherman Fairchild Foundation and Caltech.

\end{acknowledgments}

\bibliography{References}

\end{document}